\documentclass{article}
\usepackage {amssymb}
\begin {document}

\author {Albert Schwarz
\\University of California at Davis,\\Davis,CA 95616
 \thanks{Partially
supported by NSF
grant DMS-9801009}}

\title {{\bf Noncommutative supergeometry, duality and deformations.}}
\maketitle
  \large

\begin {abstract}
\large
We introduce a notion of $Q$-algebra that can be considered as a
generalization of the notion of $Q$-manifold (a supermanifold equipped
with an odd vector field obeying $\{ Q,Q\} =0$). We develop the theory 
of
connections on modules over $Q$-algebras and prove a general duality
theorem for gauge theories on such modules. This theorem containing as 
a
simplest case $SO(d,d,{\bf Z})$-duality of gauge theories on
noncommutative tori can be applied also in more complicated 
situations.
We show that $Q$-algebras appear naturally in Fedosov construction of
formal deformation of commutative algebras of functions and  that
similar $Q$-algebras can be constructed also in the case when the
deformation parameter is not formal.
\end {abstract} 
  It was shown recently that  noncommutative geometry is quite useful 
in
the study of string theory/M-theory (see [4]-[10] and references 
therein).
It
was proved ,in particular,
that the gauge theory on noncommutative tori has $SO(d,d,{\bf Z})$ 
duality
group, closely related to T-duality in string theory [5].A very general 
duality theorem, containing
$SO(d,d,{\bf Z})$-duality as a special case was derived in [11]. This 
theorem was formulated
and proved in the framework of  "noncommutative supergeometry". The 
main
idea of   noncommutative geometry is to consider every associative 
algebra
as an algebra of functions on  "noncommutative space".  Of  course,
supergeometry fits very nicely in this approach: one of the most
convenient definitions of a supermanifold is formulated in terms of 
the
algebra of functions on it.  One can say that supergeometry is
"supercommutative ${\bf Z}_2$-graded  noncommutative geometry".  

   One of important notions of supergeometry is the notion of 
$Q$-manifold
(of a manifold equipped with an odd vector field $Q$ satisfying $\{ 
Q,Q\}
=0$);see [13]. The first order differential operator $\hat {Q}$
corresponding to
$Q$ obeys $\hat {Q}^2=0$; therefore the algebra of functions on a
$Q$-manifold can be considered as a differential ${\bf Z}_2$-graded
associative algebra  and it is naturally to think that differential 
${\bf
Z}_2$-graded associative algebras are analogs of $Q$-manifolds. 
However,
in [11] we introduced another notion, the notion of $Q$-algebra, that
also can be considered as a natural generalization of $Q$-manifold and
that can be used to develop the theory of connections and to prove a
general duality theorem. Namely, one can define a $Q$-algebra as a 
${\bf
Z}_2$-graded associative algebra   $A$ equipped with an odd
derivation
$Q$ obeying $Q^2a=[\omega ,a]$; here $\omega \in A$ should satisfy
$Q =0$. (One says that a linear operator acting on graded algebra is
a derivation, if it satisfies the graded Leibniz rule.) 
Of course, in the case when $A$ is supercommutative
this definition coincides with the definition of differential algebra, 
but
if we do not assume  supercommutativity this definition is more 
general. The notion of $Q$-algebra is equivalent to the notion of CDGA-curved 
differential graded algebra-introduced in [14]. It is closely related 
to the notion of $A_\infty$-algebra .

One can define an $A_\infty$-algebra as a vector space $V$ equipped
with multilinear operations $m_i$  ; these operations should satisfy 
some
relations . ( The operations $m_i$ determine a derivation of tensor
algebra over $V$; the square of this derivation should be equal to 
zero.)
In standard definition of $A_\infty$-algebra one considers operations
$m_i$ where the number of arguments $i$ is $\geq 1$.
However, one can modify the definition  including an operation $m_0$
(if the number of arguments is equal to zero, then the operation is
simply a fixed element of $V$). Using the modified definition one
can say that $Q$-algebra is an $A_\infty$-algebra where all operations
with the number of arguments $\geq 3$ vanish. ( In standard definition
this requirement leads to differential algebras.)

We define a connection on $A$-module $E$   as an odd linear map $\nabla
:E\rightarrow E $  obeying the Leibniz rule $\nabla
(ea)=\nabla e\cdot a +(-1)^{\deg e}e\cdot Qa$; the general duality 
theorem
is formulated in terms of such connections. We analyze the relation of 
the
standard definition of connection in  noncommutative  geometry to our 
one.
It seems that many well known constructions and theorems become more
transparent in the formalism $Q$-algebras. From the other side many 
considerations of [11] are similar to arguments employed previously, 
especially in [1], [2], [3]. 

Notice, that the theory of $Q$-algebras can be generalized in the following way. We can consider
a  $Q$-algebra as ${\bf Z}_2$-graded algebra $A$ equipped with an odd derivation $Q$; then
$Q^2=\rho$ is an even derivation that is not necessarily an inner derivation. In this case we
should modify the definition of connection. Namely, to specify a connection on $A$-module $E$ we
should consider along with an odd linear operator $\nabla$ obeying the Leibniz rule an even
linear operator $\sigma$ that satisfies  
$$\sigma(ea)=(\sigma e)a=e\rho (a)$$
 Using the notation (5) we can represent this relation in the form 
$$[\sigma,\hat {a}]=\widehat {\rho (a)}.$$ For the original definition of $Q$-algebras we should
take $\sigma =\hat {\omega}.$  It is easy to verify that this definition allows us to generalize
the theory of connections presented below; the only essential modification is in definition of
the curvature where we should 
replace $\hat {\omega}$ with $\sigma$ .

The present paper contains a more detailed exposition of the results of 
the letter [11] as well as some applications of these results. In 
particular, we show that $Q$-algebras appear naturally in Fedosov construction 
of formal deformations of commutative algebras of functions and that 
similar $Q$-algebras can be constructed also in the case when the 
deformation parameter is not formal. We conjecture that these $Q$-algebras can 
be used to circumvent the problem of construction of non-formal 
deformation. (In formal case modules over deformed algebra corresponed to 
modules of appropriate $Q$-algebra equipped with zero curvature connection. 
The construction of  deformed  algebra is not known in non-formal case, 
but we hope that one can use the $Q$-algebra we constructed instead of 
this unknown algebra.)

\centerline {{\bf 1. Preliminaries.}}

When we talk about  associative algebra  $A$ we always have in mind
graded (${\bf Z}$-graded or ${\bf Z}_2$-graded) unital associative 
algebra
over ${\bf C}$.  Graded commutator is defined by the formula 
$$[a,b]=ab-(-1)^{\deg a\cdot \deg b}ba.$$
In what follows all commutators are understood as graded commutators.

   A (right) module $E$ over ${A}$ is a graded vector space
with operator of multiplication on elements of $A$ from the right;
this operation should have standard properties: $(ea)\cdot b=e\cdot 
(ab)\
\  e(a+b)=ea+eb$ etc. Grading on $E$ should be compatible with
grading on $A$ (i.e. $\deg (ea)=\deg e+\deg a$). The definition of a
left module is similar; by default our modules are right modules. 
  For every module $E$ we can construct a module $\Pi E$ changing 
the grading: $\widetilde {{\rm deg}}e={\rm deg}e=1$ 
(for ${\bf Z}_2$-graded modules the operation $\Pi$ is parity 
reversion).

  A vector space $E$ is called an $(A,B)$-bimodule
if it is a left $A$-module and a right $B$-module; we
require that $(a_1e)a_2=a_1(ea_2)$ where $a_i\in A _i,\ \  e\in {\cal
E}$.

  If  $E_1,E_2$ are $A$-modules we define an
$A$-homomorphism as a map $\varphi: E_1\rightarrow E_2$
obeying  $\varphi (xa)=\varphi (x)a$.  The graded algebra of
$A$-homomorphisms of the  $A$-module $E$ into itself
(algebra of $A$-endomorphisms) is denoted by End$_{A}E$.
If $E$ is an $(A,B)$-bimodule there exist natural
homomorphisms  $A\rightarrow {\rm End}_{B}E$  and
$B\rightarrow {\rm End}_{A}E$.

  If  $E_1$ is a right $A$-module and $E_2$ is a left
$A$-module we define $E_1\otimes _{A}E_2$ as a
vector space obtained from the standard tensor product $E_1\otimes
_{{\bf C}}E_2$ by means of identification $e_1a\otimes e_2\sim
e_1\otimes ae_2$, where $e_i\in E_i,\ \  a\in A$.

   A linear map  $E_1\otimes_{{\bf C}}E_2 \rightarrow F$ can be 
considered as $F$-valued bilinear pairing $<e_1,e_2>$; this map 
descends to $E_1\otimes _AE_2$ iff $<e_1a,e_2>=<e_1,ae_2>$.

     A (finitely generated) free module $A  ^n$ over $A$ can
be defined as the space of column vector with entries from $A$ and
with componentwise  multiplication on elements of $A$. We regard
$A ^n$ as a right module, but it can be considered also as
$(A,A )$-bimodule.  (We already used the structure of
$(A,A )$-bimodule on $A ^1=A$.) The algebra
End$_{A}A^n$ of endomorphisms of $A ^n$ can be identified
the algebra Mat$_n{A}$  of  $n\times n$ matrices with entries from
$A$; these matrices act on $A ^n$ by means of multiplication
from the left. A projective $A$-module can be defined as a direct
summand $E$ in a free module $A ^n$.  The decomposition $A
^n=E+E^{\prime}$ into a direct sum determines an endomorphism $e: A
^n\rightarrow A ^n$ projecting  $A ^n$ onto $E$; in other words
$e^2=e,\ \  ex=x$ for $x\in E,\ \  ex^{\prime}=0$ for $x^{\prime}\in
E^{\prime}$.  Notice that in our  terminology projective modules are
always finitely generated.

  Projective $A$-modules form a semigroup with respect to direct
summation. Applying the Grothendieck construction to this semigroup we
obtain the K-theory group $K_0(A)$. More precisely, we say that a
projective module $E$ specifies an element $[E]\in K_0(A)$  and
impose the relations $[E _1+E_2]=[E_1]+[E_2],\ \ [E+\Pi E]=0$. If we 
work with 
${\bf Z}_2$-graded modules there is no necessity to consider formal 
differences $E_1-E_2$ (virtual modules); the relation $[E+\Pi E]=0$ 
permits us to replace virtual module $E_1-E_2$ with ${\bf Z}_2$-graded 
module $E_1+\Pi E_2$. 
A ${\bf C}$-linear map
$\tau : A\rightarrow {\bf C}$ is called a (graded ) trace if it
vanishes on all (graded) commutators: $\tau ([a,b])=0$ for all
$a,b\in A$. We always consider graded traces; therefore we almost
always omit the word "graded"  in our formulations. 

  A trace $\tau$ on  $ A $ generates a trace on End$_{A}A
^n=$Mat$_n  A$; this trace will be denoted by the same symbol $\tau$.
(To calculate the trace of a matrix $(a_{ij})\in$Mat$_n A$ one should
take the supertrace of the matrix $(\tau(a_{ij}))$. 

  If  $E\subset A ^n$ is a projective module then the algebra
End$_{A}E$ of endomorphisms of $E$ can be identified with the
subalgebra of End$_{A}A ^n=$Mat$_n\Omega$ consisting of elements
of the form $eae$. (Here $e:A ^n\rightarrow A ^n$ is a
projection of $A^n$ onto $E,\ \  a\in $End$_{A}A ^n$). We
define a graded trace $\tilde {\tau}$ on  End$_{A}E$ as a restriction
of $\tau$ to this subalgebra. 

     If $E$ is an $A$-module, then starting with an element
$g\in E$ and $A$-homomorphism $f: E\rightarrow A$
we can construct an endomorphism $g\otimes f:E\rightarrow E$
transforming $x\in E$ into $gf(x)\in E$. (The endomorphism
$g\otimes f$ can be considered as a generalization of linear operator 
of
rank $1$.)

   For any algebra $A $ we construct a vector space $\bar {A}=
A /[ A , A]$ factorizing the vector space $ A $ with
respect to the subspace $[ A , A ]$ spanned by all (graded)
commutators $[a,b]$.  This construction is closely related with the 
notion
of trace: traces on  $A$  correspond to linear functionals on  $\bar
{A }$.  

   If $E$ is a projective $A $-module, one can construct a
${\bf C}$-linear map Tr: End$_{A}E\rightarrow \bar {A}$
transforming an endomorphism of the form $g\otimes f$ into the class
$\overline {f(g)} \in \bar {A}$ of  $f(g)\in {A}$. (Such a ${\bf
C}$-linear map is unique because in the case of projective module 
every
endomorphism can be represented as a finite sum of endomorphisms of 
the
form $g\otimes f$.) The map Tr has the main property of trace 
$${\rm Tr} [\varphi ,\psi ]=0$$
(trace of graded commutator of two $A$-endomorphisms $\varphi,\psi
\in {\rm End}_{A}E$ vanishes).  In some sense the map
Tr:End$_{A}E\rightarrow \bar {A}$ can be considered as
universal trace on End$_{A}E$. (As we mentioned every trace
$\tau$ on $A$ determines a trace $\tilde {\tau}$ on
End$_{A}E$. It is easy to verify that $\tilde
{\tau}(\varphi)=\tau({\rm Tr }\varphi)$.)

\centerline {{\bf 2. $Q$-algebras.}}

{\bf Definition.} {\it  Let $A$ be a graded associative algebra. We say the
$A$ is a $Q$-algebra if it is equipped with derivation $Q$ of degree
$1$ and there exists an element $\omega \in A ^2$ satisfying 
\begin {equation}
Q ^2x=[\omega ,x]
\end {equation}
for all $x\in A$.}

  Calculating $Q^3x$ in two ways we obtain 
$$Q^3x=Q([\omega,x])=Q\omega\cdot x +\omega\cdot Qx-Qx\cdot \omega 
-x\cdot
Q\omega\cdot (-1)^{\deg x}$$

$$Q^3x =Q^2\cdot Qx=[\omega, Qx]=\omega\cdot Qx-Qx\cdot \omega.$$
We see that $Q\omega\cdot x=x\cdot Q\omega\cdot (-1)^{\ deg x}$, i.e. 
\begin {equation}
[Q\omega,x]=0
\end {equation}
We proved that $Q\omega\in A ^3$ commutes with all elements of
$A $ (in the sense of superalgebra). In almost all interesting cases
it follows from this condition that $Q\omega $ vanishes. 

{\it We will include the additional condition 
\begin {equation}
Q\omega =0
\end {equation}
in the definition of $Q$-algebra.}

We almost always consider unital algebras. It is easy to to check that 
applying
$Q$ to the unit we get 0. (This follows from the  Leibniz rule.)

   Let us consider a (graded) $A$-module $E$. We define a connection
on $E$ as a ${\bf  C}$-linear operator $\nabla :E\rightarrow 
E$ having degree $1$ and obeying the Leibniz rule:
\begin {equation}
\nabla (xa)=(\nabla x)\cdot a+(-1)^{\deg x}\cdot x\cdot Qa.
\end {equation}
for all $x\in E,\ \ a\in A$.

  Let us introduce the notation 
\begin {equation}
 \hat {a}x=(-1)^{\deg x\cdot \deg a}xa
\end {equation}
The formula (4) can be rewritten in the form 
\begin {equation}
[\nabla,\hat {a}]=\widehat {Qa}
\end {equation}
It is easy to check that some standard statements about connections 
remain
true in our case. However, the definition of curvature should be 
modified.

  1) If  $\nabla $ is a fixed connection on $E$, then every other
connection has the form 
$$\nabla ^{\prime}=\nabla +A$$
where $A\in {\rm End}^1_{A}E$ is an arbitrary endomorphism of
degree $1$.

  2)  If $\varphi \in{\rm End}_{A}E$ is an endomorphism then
$[\nabla ,\varphi ]$ is also endomorphism. 

  3) The operator $\nabla ^2+\hat {\omega}$ is an endomorphism: 
$\nabla
^2+\hat {\omega} \in {\rm End}^2_{A}E$. This endomorphism is
called the curvature of connection $\nabla$; it is denoted by 
$F(\nabla)$
(or simply by $F$). It obeys $[\nabla, F]=0$. 

  To check this statement we represent $\nabla ^2$ as ${1\over
2}[\nabla,\nabla]$ and calculate $[[\nabla,\nabla],\hat {a}]$ by means 
of
(4) and Jacobi identify.

  4) Let us define the operator $\tilde {Q}: {\rm End}_{A}
E\rightarrow {\rm End}_{A}E$ by the formula 
$$ \tilde {Q} \varphi =[\nabla,\varphi].$$
It is easy to verify that 
\begin {equation}
 \tilde {Q}^2\varphi =[F,\varphi],
\end {equation}
where $F$ is the curvature of $\nabla$
It follows from this statement and from $\tilde {Q}F=[\nabla,F]=0$ 
that
the algebra ${\rm End}_{A}E$ equipped with the operator
$\tilde {Q}$ is a $Q$-algebra with $\tilde {\omega}=F$. (One should
notice, however, that we can also take $\tilde {\omega}=F+c$, where $c$ 
is
a central element obeying $\nabla c=0$.)

 5) If $E$ is considered as a module over $Q$-algebra ${\rm End}_AE$, 
then $\nabla$ is a connection on this module.

  As we mentioned, the notion of  $Q$-algebra is a generalization of 
the notion of differential algebra. It is important to notice that the 
endomorphism algebra End$_AE$ is not  necessarity a  differential 
algebra: we have $\tilde{Q}^2=0$ only in the case when $F$ is a central 
element. In particular, a structure of differential algebra on ${\rm End}_AE$ arises 
if  $\tilde {Q}$ is defined by means of constant curvature connection 
$\nabla$.

\centerline {{\bf 3. Equivalent $Q$-algebras.}}

Let us consider a graded associative algebra $A$. Let us suppose that a 
structure of $Q$-algebra on $A$ 
is specified by means of operator $Q$ obeying (1). If $\gamma$ is 
element of $A$ we denote by $\tilde{\gamma}$ a derivation of $A$ defined by 
the formula 
\begin {equation}
\tilde {\gamma}(a)=[\gamma ,a].
\end {equation}
Taking $\gamma \in A^1$ we can construct a derivation of degree 1:
\begin {equation}
Q'=Q+\tilde {\gamma}.
\end {equation}
It is easy to check that 
\begin {equation}
{Q'}^2x=[\omega ',x]
\end {equation}
where $\omega '=\omega +(Q\gamma +\gamma ^2)$ and $Q'\omega '=0.$

This means that $Q'$ specifies another  structure of $Q$-algebra on
 $A$. We will show that this new structure is in some sense equivalent
to the  original one. More precisely, we fix an $A$-module $E$ and 
consider connections on $E$ with respect to the original $Q$-structure
 ($Q$-connections) and with respect to new $Q$-structure  
($Q'$-connections). We will prove that {\it there exists 
one-to-correspondence between   $Q$-connections and $Q'$-connections.}
 Namely, for every  $Q$-connection $\nabla : E\rightarrow E$ we can 
construct a $Q'$-connection  $\nabla '=\nabla-\hat {\gamma}$, where 
$\hat {\gamma}$ is defined by (5). (To check that $\nabla '$ is a
 $Q'$-connection we use (6).) {\it The curvature $F'$ of $\nabla '$ is 
 equal  to  the curvature} $F$ of $\nabla $ .

As we mentioned in Sec.2 a $Q$-connection $\nabla :E\rightarrow E$ induces a structure of
$Q$-algebra on the algebra of endomorphisms ${\rm End}_AE$; the perator $\tilde {Q}$ on ${\rm
End}_AE$ is defined by the formula $\tilde {Q}_{\varphi}=[\nabla ,\varphi]$.  It is obvious that
replacing a connection $\nabla$ with another connection $\nabla +\alpha$ we obtain an equivalent
$Q$-algebra structure specified by the operator transforming $\varphi\in {\rm End}_AE$ into
$\tilde {Q}\varphi +[\alpha, \varphi ]$. Identifying equivalent $Q$-structures we can say that
$Q$-structure on ${\rm End}_AE$ does not depend on the choice of connection $\nabla$. 

Let us define a $dg$-module over a $Q$-algebra as a module equipped 
with a connection with $F=0$ (zero curvature connection). This terminology 
agrees with standard terminology in the case when a $Q$-algebra is a 
$dg$-algebra ($Q^2=0$), because in this $F=\nabla ^2$ and $\nabla$ can be 
regarded as a differential. It follows from the above statements that 
there exists a one-to-correspondence between $dg$-modules over 
$Q$-algebra $(A,Q)$ and $dg$-modules over equivalent $Q$-algebra 
$(A,Q')=(A,Q+\tilde {\gamma})$. If $Q'^2=0$ (a $Q$-algebra is equivalent to a 
differential algebra), we can reduce the study of $dg$-modules over $Q$-algebra 
to the study of $dg$-modules over equivalent  differential algebra (or 
to go in opposite direction if the  differential algebra is more 
complicated.)
  
\centerline {{\bf 4. Connections on projective modules.}} 

   First of all it is easy to construct a connection on an arbitrary
projective $A$-module $E$  where $A$ is a $Q$-algebra. Namely, if $E$ 
is specified by means of projection $e:A^n\rightarrow A^n$ (i.e. 
$e\Omega^n=E$) we can construct a connection on $E$ (so called
Levi-Civita connection)  by means of the formula $D=eQe$ where $Q$ acts 
on
$A ^n$ componentwise. (The Leibniz rule for $D$ follows from $e^2=e$
and from the Leibniz rule for $Q$.) The curvature of the Levi-Civita
connection is given by the formula: 
$$F=e((Qe)^2+\omega\cdot 1).$$ 
 For any algebra $A$ we defined a vector space
$\bar{A}=A/[A,A]$. If $A$ is a $Q$-algebra we
have $Q([A,A])\subset [A,A]$. This means that the
operator $Q: A\rightarrow A $ descends to an operator $\bar
{Q}:\bar {A}\rightarrow \bar {A}$.  It is easy to check that
$\bar {Q}$ is a differential: $\bar {Q}^2=0$.

   Now we will define the Chern character of a connection $D$ on a
projective $A$-module $E$ as an element of $\bar {A}$:
$${\rm ch}D=\sum_{q=0}{1\over q!}{\rm Tr}F^q$$
(Recall that we defined a map ${\rm Tr: End}_{A}E\rightarrow
\bar {A}$ using the formula ${\rm Tr} (g\otimes f)=\overline {f(g)}$.
Here $f: E \rightarrow A$ is an $A$-homomorphism, $g\in
A$ and $g\otimes f$ transforms $x\in E$ into $gf(x)\in {\cal
E}$. The map $a\rightarrow \bar {a}$ transforms $a\in A$ into its
class $\bar {a}\in \bar {A}$.)

   One can prove the following statements: 

1) ${\rm ch}D$ is closed with respect to the differential $\bar {Q}$ 
in
$\bar {A}$:
\begin {equation}
\bar {Q}{\rm ch}D=0
\end {equation}

2) If $D^{\prime},D$ are two connections on $A$-module $E$
then ${\rm ch}D^{\prime}-{\rm ch}D$ is exact with respect to the
differential $\bar {Q}$: 
\begin {equation}
{\rm ch}D^{\prime}-{\rm ch}D=\bar {Q}{\rm (something)}.
\end {equation}
The proof is based on the following lemma:

For every endomorphism $\varphi \in {\rm End}_{A}E$ we have 
\begin {equation}
{\rm Tr} [D,\varphi]=\bar {Q}{\rm Tr}\varphi
\end {equation}
It is sufficient to verify (14) for Levi-Civita connection $D=eQe$
(because ${\rm Tr} [D^{\prime}-D,\varphi]=0$) and for endomorphisms of 
the
form $\varphi =g\otimes f $ (because these endomorphisms span
End$_{A}E$).

  Using (14) we deduce (12) from the relation $[D,F^q]=0$ that follows
immediately from $[D,F]=0$.
 
  To derive (13) we will consider a smooth family 
$D(t)=D+t(D^{\prime}-D)$
of connections on $E$ and prove that 
$${d\over dt}{\rm ch}D(t)=\bar{Q}({\rm something}).$$
First of all we notice that the curvature $F(t)$ of connection $D(t)$
obeys 
$${dF(t)\over dt}=[\Gamma,D(t)]$$ 
where $\Gamma =D^{\prime}-D\in {\rm End}_{A}E$. We see that 
$${dF\over dt}=[\Gamma,D] \ \ \ {\rm mod} [\bar {A},\bar {A}].$$
and therefore
$${dF^q\over dt}=q[\Gamma,D]F^{q-1}=q[D,\Gamma F^{q-1}] \ \ \  {\rm
mod}[\bar {A},\bar {A}],$$
$${d{\rm Tr}F^q\over dt}=q{\rm Tr}[D,\Gamma F^{q-1}]\in \bar {Q}(\bar
{A}).$$
Integrating over $t$ we obtain (13). 

   In the proof of (13) we assumed that $A $ is equipped with
topology having some properties that permit us to justify the 
calculations
above. These assumptions are not necessary; it is easy to modify our
consideration  to obtain completely algebraic proof (as in [2] for
example). 

   Sometimes it is convenient to reformulate (13) using the notion of
closed trace. We say that a linear functional on $A$ is a closed
trace if it vanishes on (graded) commutators and on elements of the 
form
$Qa$.  It follows from (13) that for a closed trace $\tau$ the number
$\tau({\rm ch}(D))$ does not depend on the choice of the connection $D$ 
on
the module $E$; it depends only on the $K$-theory class of the
module $E$. 

   Using the differential $\bar {Q}$ we can define the homology 
$H(\bar
{A})$ in the standard way: $H(\bar {A})=\ker \bar {Q}/{\rm
Im}\bar {Q}$. It follows from (12), (13) that the Chern character
specifies a homomorphism ch: $K_0(A)\rightarrow H^{{\rm
even}}(\bar{A})$.

\centerline {{\bf 5. Morita equivalence of $Q$-algebras}}

  Let us consider an $(A,B)$-bimodule $P$ where $A$ 
is a $Q$-algebra with respect to the operator $Q_1$ and $B$ is
a $Q$-algebra with respect to the operator $Q_2$. We say that an 
operator
$\nabla _P:P\rightarrow P$ is a connection on bimodule $P$ if 
$$\nabla _P(ax)=(-1)^{\deg a}\cdot a \nabla _P(x)+Q_1a\cdot x$$
$$\nabla _P(xb)=\nabla_Px\cdot b+(-1)^{\deg x}\cdot x Q_2b$$
for all $x\in P,\ \ a\in A _1,\ \  b\in B_2$.

   In other words, $\nabla _P$ should be a connection with respect to
$A$ and with respect to $B$ at the same time. 

  It follows from the above statements that every $A$-module $E$  equipped with a connection
$\nabla$ can be considered as $({\rm
End}_{A}E,A)$-bimodule and $\nabla$ is a connection on
this bimodule.

  Using an $(A, B)$-bimodule $P$ we can assign to every
(right ) $A$-module $E$ a (right) $B$-module
$\tilde {E}$ taking the tensor product with $P$:
\begin {equation}
\tilde {E}=E\otimes _{A}P
\end {equation}
(To take the tensor product over $A$ we identify $ea\otimes p$
with $e\otimes ap$ in the standard tensor product $E\otimes _{\bf
C}P$. Here $e\in E,\ \  p\in P,\ \  a\in A$.)

   If we have a connection $\nabla _P$ in the bimodule $P$ we can 
transfer
a connection on $E$ to a connection on $\tilde {E}$. Namely,
for every connection $\nabla $ on $E$ we define an operator $\nabla
\otimes 1+1 \otimes \nabla_P$ on $E\otimes _{\bf C}P$. It is easy
to check that this operator is compatible with identification 
$ea\otimes
p\sim e \otimes ap$ and therefore descends to an operator $\tilde
{\nabla}: \tilde {E} \rightarrow \tilde {E}$. The operator
$\tilde {\nabla}$ can be considered as a connection on $B$-module
$\tilde {E}$. 
  
   It is easy to relate the curvatures of the connections $\nabla$ and
$\tilde {\nabla}$. We should take into account that correspondence 
between
$E$ and $\tilde {E}$ is natural, i.e. to every endomorphism
$\sigma \in  {\rm End}_{A}E$ we can assign an endomorphism
$\tilde {\sigma}\in {\rm End}_{B} \tilde {E}$ (the map
$\sigma \otimes 1:E\otimes _{{\bf C}}P\rightarrow E\otimes
_{{\bf C}}P$ descends to an endomorphism $\tilde {\sigma}: \tilde {
E}\rightarrow \tilde {E}$). In particular, the curvature
$F(\nabla)
\in {\rm End}_{A}E$ determines an endomorphism $\widetilde
{F(\nabla)} 
\in  {\rm End}_{B} \tilde {E}$. One can verify that the
curvature $F(\tilde {\nabla})$ of the connection $\tilde {\nabla}$ on
$\tilde {E}$ can be represented in the form: 
\begin {equation}
F(\tilde {\nabla})=\widetilde {F(\nabla)}+\tilde {\varphi},
\end {equation}  
where $\tilde {\varphi}$ is a fixed element of  ${\rm End}_{B}
\tilde {E}$ .

  To verify (7) we notice that 
$$\nabla ^2\otimes 1+\hat {\omega}_1\otimes 1: E\otimes _{{\bf
C}}P\rightarrow E\otimes _{{\bf C}}P$$ descends to the endomorphism
$\widetilde {F(\nabla)}: \tilde {E}\rightarrow \tilde {E}$
and $\nabla ^2\otimes 1+1\otimes \nabla^2_P+1\otimes \hat {\omega}_2$
descends to  $F(\tilde {\nabla}): \tilde {E}\rightarrow \tilde
{E}$. Using the relation $\hat {\omega }_1\otimes 1=-1\otimes
\hat{\omega}_1$ we obtain that the map $\tilde
{\varphi}=F(\tilde{\nabla})-\widetilde{F(\nabla)} $ is induced by the 
map
$\varphi =1\otimes \psi : E\otimes _{{\bf  C}}P\rightarrow 
E\otimes _{{\bf C}}P$ where the map $\psi : P\rightarrow P$ is given 
by
the formula 
$$\psi =\nabla ^2_P+\hat {\omega}_1+\hat {\omega }_2.$$
It is easy to check  that
\begin {equation}
 \psi  \in {\rm End}_{A} P\cap {\rm End}_{B} P 
\end {equation}
(i.e. $\psi (ax)=a\psi (x),\ \  \psi (xb)=\psi (x)b$ for $x\in P,\ \  
a\in
A,\ \  b\in B$). To check that $\psi$ commutes with $a\in
A$ we represent it in the form $\psi =\widehat
{F_1(\nabla_P)}+\hat {\omega}_2$, where $F_1(\nabla_P)$ stands for the
curvature of $\nabla_P$ considered as $A$-connection; the
representation $\psi =\widehat {F_2(\nabla_P)}+\hat {\omega}_1$ should 
be
used to prove that $\psi \in {\rm End }_{B}P $. 

   It follows from (8) that $\varphi =1\otimes \psi $ descends to 
$\tilde
{E}$ and gives an $B$-endomorphism $\tilde {\varphi}$. One
should notice that these facts are clear also from the representation
$\tilde {\varphi}=F(\tilde {\nabla})-\widetilde {F(\nabla)}$. 

   To illustrate the above statements we can start with an arbitrary
$Q$-algebra $A$ and arbitrary $A$-module $P$ with connection
$\nabla _P$. We consider $P$ as $(A,B)$-bimodule, where
$A={\rm End }_{A}P,\ \  B=A$. (We have seen that
$A={\rm End}_{A}P$ is a $Q$-algebra with respect to the
operator $\tilde {Q}\varphi =[\nabla _P,\varphi ]$ and that $\nabla _P$ 
is
a connection also with respect to this $Q$-algebra.) It follows from 
our
calculations that $F=F_2(\nabla_P)=\nabla ^2_P+\hat {\omega}_2,\ \  
\hat
{\omega }_1=-F$ and therefore $\psi =F+\hat {\omega}_1=0$. (We can 
obtain
the same result noticing that $F_1(\nabla
_2)=\nabla^2_P+\hat{\omega}_1=(F-\hat {\omega}_1=-\hat {\omega}_2$.) 
We
see that in our situation $\varphi =0$;  hence, $F(\tilde
{\nabla})=\widetilde {F(\nabla)}$. (However, as we noticed above one 
can
modify the definition of $Q$-algebra ${\rm End} _{A}P$ adding central
element $c$ with $\nabla c=0$ to $\omega _1$; then $\varphi \not= 0$.)

   We would like to give conditions when gauge theories on
$A$-module $E $ and in $B$ -module $\tilde {
E}$ are equivalent. To establish such an equivalence we need $(A, 
B)$-bimodule $P^{\prime}$ equipped with connection
$\nabla_{P^{\prime}}$. Such a bimodule permits us to transfer modules 
and
connections in opposite direction. If the constructions obtained by 
means
of $P^{\prime}$ are inverse to constructions specified by $P$ we say 
that
bimodules $P,P^{\prime}$ give  Morita equivalence of $Q$-algebras
$A$ and $B$ (or that they are  Morita equivalence
bimodules). Of course, this notion generalizes the standard notion of
Morita equivalence of associative algebras, when we do not use the
operator $Q$ and connections. The definition of Morita equivalence
bimodules can be reformulated in the following more constructive way. 
Let
us suppose that there exist two bilinear scalar products between $P$ 
and
$P^{\prime}$ taking values in $A$ and in $B$ respectively.
We assume that scalar products are $A$-invariant and $B$-invariant 
correspondingly. In other words, we assume that for $p\in
P,\ \  p^{\prime}\in P^{\prime}$ we have $<p,p^{\prime}>_1\in A,\ \
<p^{\prime},p>_2\in B$ and $<p\omega,p^{\prime}>_1=<p,\omega
p^{\prime}>_1$ for $\omega \in B$,
$<p^{\prime}\sigma_1,p>_2=<p^{\prime},\sigma_1 p>_2$ for $\sigma_1 \in
A$. We require also that 
\begin {equation}
\sigma_1 <p,p^{\prime}>_1\sigma_2=<\sigma _1p,p^{\prime}\sigma _2>_1,\ 
\
\omega_1<p^{\prime},p>_2\omega _2=<\omega_1p^{\prime},p\omega_2>_2
\end {equation}
\begin {equation}
p_1<p,p^{\prime}>_1=<p_1,p>_2p^{\prime},\ \
<p^{\prime},p>_2p_1^{\prime}=p^{\prime}<p,p_1^{\prime}>_1
\end {equation}
Here $p,p_1\in P,\ \  p^{\prime},p_1^{\prime}\in P^{\prime},\ \  
\sigma
_i\in A,\ \  \omega _i\in B$. The scalar products
determine maps 
$$\alpha :P\otimes _{B}P^{\prime}\rightarrow A,\ \  \beta
:P^{\prime} \otimes _{A} P \rightarrow B.$$
We can consider $P\otimes _{B}P^{\prime}$ and $A$ as 
   $(A,A)$-bimodules; then it follows from (9), that
$\alpha$ is a homomorphism of bimodules; similarly $\beta$ is a
homomorphism of   $(B,B)$-bimodules. We require that
$\alpha$ and $\beta$ be isomorphisms. Then 
$$(E\otimes _{A} P)\otimes _{B}P^{\prime}=
E\otimes _{A}(P\otimes_{B}P^{\prime})=E\otimes
_{A}A=E$$
for every  $A$-module $E$. This statement together with
similar statement for   $B$-modules gives us one-to-one
correspondence between    $A$-modules and    $B$-modules
(more precisely it gives us equivalence of categories of  $A$-modules
 and    $B$-modules).  To obtain  one-to-on
correspondence between  connections we should impose additional
requirements 
\begin {equation}
<\nabla _Pp,p^{\prime}>_1+<p,\nabla_{P^{\prime}}p^{\prime}>_1=
Q<p, p^{\prime}>_1, 
\end {equation}
$$<\nabla_{P^{\prime}}p^{\prime},p>_2+<p^{\prime},\nabla_{P^{\prime}}p>_2=
Q<p^{\prime}, p>_2$$

It follows from our assumptions that the operator  $$\nabla _P \otimes 
1 +
1 \otimes \nabla _{P^{\prime}}$$ on $P \otimes _{\bf C} {P^{\prime}}$
descends to operator $Q$ on
$P \otimes _{B} {P^{\prime}}$. Using that $Q\cdot  1 =0$  we obtain  
that the operator $\nabla \otimes 1 + 1 \otimes Q$ on
$E \otimes_{\bf C}A$ descends to $\nabla$ on
$E \otimes_{A}A=E$. This means that going from
$A$-connection to $B$-connection and back we obtain the
original $A$-connection. This fact together with similar statement
about $B$-connections gives one-to-one correspondence between 
$A$-connections and $B$-connections.

We see that {\it under our
 conditions we have equivalence between gauge theories on 
  $A$-module $E$ and on   $B$-module $\tilde {
E}$} (duality). We will describe later how the duality of gauge 
theories
on noncommutative tori can be obtained this way. 

 Using well known results about Morita equivalence associative algebras [12] 
one can describe $Q$-algebras that are equivalent to a given $Q$-algebra 
$A$ in the following way. Let us consider a projective $A$-module $P$ 
that is equipped with a connection $\nabla $. Let us assume that $P$ is a 
generator ( i.e. $A^1$ is a direct summand in $P^n$). Then $\hat 
{A}={\rm End}_A P$ is   Morita equivalent to $A$ as a $Q$-algebra. (The 
structure of $Q$-algebra on $\hat {A}$ is specified by an operator $\hat 
{Q}$ defined by the formula $\hat {Q}\varphi =[\nabla, \varphi]$.) {\it All 
$Q$-algebras that are Morita equivalent to $A$ can be obtained by means of 
this construction.} (This follows easily from the remark that two  
Morita equivalent $Q$-algebras are  Morita equivalent as  associative 
algebras.)

\centerline {{\bf  6. Connections on modules over associative 
algebras.}}

   The theory of connections on modules over $Q$-algebras can be
considered as a generalization of the theory of connections on 
associative
algebras. If  $A$ is an associative algebra one can construct a
differential ${\bf Z}$-graded algebra $\Omega (A)=\sum _{n\geq 
0}\Omega
^n(A)$ (universal differential graded algebra) in the following way. 
The
vector space $\Omega ^n(A)$ is is spanned by formal expressions
$a_0da_1...da_n$ and $\lambda da_1...da_m$ where $a_0,...a_n\in A,\ \
n\geq 0,\ \  m\geq 1,\ \  \lambda \in {\bf C}$. The multiplication and 
the
differential on $\Omega (A)$ are defined by means of Leibniz rule and
relation $d^2=0$. If $E$ is an $A$-module we define an $\Omega
(A)$-module ${\cal E}$ as a tensor product:$
{\cal E}=E\otimes _A\Omega (A)$ where $\Omega (A)$ is considered
as $(A,\Omega (A))$-bimodule. We can define a connection on $A$-module
$E$ as a connection of $\Omega (A)$-module ${\cal E}$;
this definition is equivalent to the definition given by Connes 
(see[1]).

  Closed traces on $\Omega (A)$ can be identified with cyclic cocycles 
of  
algebra; taking into account that for closed trace $\tau$ the number 
$\tau ({\rm ch}D)$ does not depend on the choice of connection $D$ we 
obtain a pairing of $K_0(E)$ with cyclic cohomology.

   In the definition of connection on $A$-module $E$ the 
algebra $\Omega(A)$ 
can be replaced with any differential extension of the algebra $A$ 
(with
any
differential graded algebra $\Omega$ that contains $A$ as a subalgebra
of
$\Omega ^0$). Moreover, one can consider any $Q$-extension of $A$ (any
$Q$-algebra $\Omega$ obeying $A\subset \Omega^0$) and define a
connection on $A$-module $E$ as a connection on $\Omega$-module
$E\otimes _A\Omega$. It is interesting to notice that under certain
conditions on algebra $\Omega $ any projective $\Omega$-module ${\cal  
E} $ can be represented in the form $E\otimes _A\Omega$ where $E$ is
projective $A$-module, $A=\Omega^0$ (see[12]). In particular, this
statement is correct if $\Omega=\sum_{0\leq k\leq n}\Omega^k$, i.e. the 
degree of an element of $\Omega$ is non-negative and bounded from
above. (In this case elements of $\Omega$  having positive degree form 
a nilpotent ideal $I$ and $A$ can be identified with $\Omega /I$.) 

   If a Lie algebra $L$ acts on $A$ by means of
infinitesimal automorphisms  (derivations) we can construct a 
differential
graded algebra $\Omega=\Omega (L,A)$ of cochains of Lie algebra $L$ 
with
values  in $A$. The elements of $\Omega$ can be considered as 
$A$-valued
functions of anticommuting variables $c^1,....,c^n$ corresponding to 
the
elements of the basis $\delta_1,...,\delta_n \in L$;  the differential 
$d$
has the form 
$$d\omega =(\delta
_{\alpha}\omega)c^{\alpha}+{1\over2}f^{\alpha}_{\beta\gamma}c^{\beta}c^{\gamma}{\partial\over\partial
c^{\alpha}}$$ 
where $f^{\alpha}_{\beta\gamma}$ are the structure constants of $L$ in 
the
basis $\delta_1,...,\delta _n$. 

  In other words we can describe the vector space $\Omega (L,A)$ as a
tensor product $\Lambda (L^*)\otimes A$ where $L^*$ stands the vector
space dual to $L$ and $\Lambda (M)$ denotes the Grassmann algebra
generated by vector space $M$ (as vector space $\Lambda (M)$ is a 
direct
sum of antisymmetric tensor powers of $M$). The grading on $\Omega 
(L,A)$
is defined by means of the natural grading on $\Lambda L^*$; if $A$ is 
a
graded algebra one should take into account also the grading on $A$.

  Let us consider in more detail connections on $A$-module $E$ with
respect
to differential extension $\Omega =\Omega(L,A)$, i. e. connections on 
graded $\Omega$-module ${\cal
E}=E\otimes _A\Omega$. In this case ${\cal
E}^0=E\otimes _A\Omega^0=E,\ \  {\cal E}^1=E\otimes _A\Omega^1=E\otimes
_{{\bf C}} L^*$. The elements $e\otimes \omega,\ \  e\in E,\ \  
\omega\in
\Omega$ span $E$, therefore, the connection $\nabla :{\cal
E}^r\rightarrow {\cal E}^{r+1}$ is completely determined by the map
$\nabla :E^0\rightarrow E^1$ that can be considered as a map
$\nabla :E\rightarrow E\otimes L^*$ or as a family of maps
$\nabla_x:E\rightarrow E$ that depend linearly on $x\in L$. Instead of 
the
family  $\nabla _x$ we can consider $n$ maps $\nabla _1,...,\nabla_n$
corresponding to the elements of the basis $f_1,...,f_n$ of the Lie
algebra $L$. These maps obey the Leibniz rule 
$$\nabla _{\alpha}(ea)=\nabla _{\alpha}e\cdot a+e\delta _{\alpha}a$$
where $\delta _{\alpha}$ stands for the derivation of the algebra $A$
that corresponds to $f_{\alpha}\in L$.

   Notice that  the Grassmann algebra  $\Lambda (L^*)$ is 
supercommutative, therefore  the space $\bar {\Omega }=\Omega 
/[\Omega ,\Omega ]$ that corresponds to $\Omega=\Omega (L,A)= 
\Lambda {L^*}\otimes A$ can be identified with $\Lambda (L^*) 
\otimes \bar {A}$. This means that the Chern character takes values 
in $H^{{\rm even}}(\bar {\Omega})=H^{{\rm even}}(L,\bar {A})$ (in 
the cohomology of the Lie algebra $L$ with coefficients in $\bar 
{A}$).

 Let us consider an $(A,\hat {A})$-bimodule $P$ assuming that Lie 
algebra $L$ 
acts on $A$ and  $\hat {A}$ by means of infinitesimal automorphisms. We 
would 
like to transfer connections from $A$-module $E$ to $\hat {A}$-module 
$\hat {E}=E\otimes _AP$. It is easy to see that we can do this using 
the formula 
$\hat {\nabla}_x=\nabla _x\otimes 1+1\otimes \nabla_x^P$ if the 
bimodule $P$ 
equipped with constant curvature connection $\nabla _x^P $ (i.e.  we 
have family 
of maps $\nabla _x^P$ satisfying the Leibniz rule with respect to $A$ 
and 
$\hat {A}$, the curvature $F_{xy}=[\nabla_x,\nabla_y]-\nabla_{[x,y]}$ 
should be 
equal to $f_{xy}\cdot 1$). If $P$ generates Morita equivalence between 
$A$ and $\hat {A}$ and the map $\nabla\rightarrow \hat {\nabla}$ is one-to-one correspondence
between connections on $E$ and $\hat {E}$ we say that $A$ and 
$\hat {A}$ are gauge Morita equivalent (in [5]  I  used the term 
"complete  Morita 
equivalence" for this notion).
  It is easy to check that endomorphism algebra End$_{\Omega (L,A)}{\cal P}$ of the  
$\Omega (L,A)$-module ${\cal P}=P\otimes_A \Omega (L,A)$ is isomorphic 
to $\Omega (L,\hat {A})$. (If  we consider elements of ${\cal P}$ as 
$P$-valued functions of anticommuting variables $c^1,...,c^n$  then   multiplying an element 
of ${\cal P}$ from the left by an $\hat {A}$-valued function of 
$c^1,...,c^n$ we 
obtain an endomorphism of ${\cal P}$. The homomorphism from 
$\Omega (L,\hat {A})$ into End$_{\Omega (L,A)}{\cal P}$ obtained this 
way is 
an isomorphism.) It follows from this remark that   $\Omega (L,\hat 
{A})$ is Morita 
equivalent to $\Omega (L,A)$. A connection on $P$ is by definition a 
connection on 
${\cal P}$; if the connection has a constant curvature than 
corresponding operator 
$\tilde {Q}$ on  $\Omega (L,\hat {A})={\rm End}_{\Omega (L,\hat 
{A})}{\cal P}$ 
is a differential. It is easy to check that under the conditions above 
$\tilde {Q}$ 
coincides with the differential on  $ \Omega (L,\hat {A})$ considered 
as algebra of 
cochains of Lie algebra $L$ with values in  $\hat {A}$. We obtain that 
differential 
algebras  $\Omega (L,A)$ and  $ \Omega (L,\hat {A})$ are Morita 
equivalent as 
$Q$-algebras if the algebras $A$  and $\hat {A}$ are gauge Morita 
equivalent. 
  
  Let $A$ be an algebra $A_{\theta}$ of smooth functions on
$d$-dimensional noncommutative torus (i.e. an algebra of expressions 
of
the form $\sum c_nU_n$, where $c_n$ is a ${\bf C}$-valued function on 
a
$d$-dimensional lattice that vanishes at infinity faster than any 
power
and the multiplication is defined by the formula $U_nU_m=\exp (\pi 
i\theta 
_{nm})U_{n+m}$, where $\theta _{nm}$ is a bilinear function on the 
lattice). 
Then it is natural to construct a differential extension 
$\Omega_{\theta}$
of $A_{\theta}$ taking as $L$ the Lie algebra of derivations $\delta 
_x$
where $\delta _xU_l=<x,l>U_l$. (We assume that the lattice is embedded
into vector space $V$. The vector $x$ belongs to the dual space $V^*$ 
that
can be identified with the Lie algebra $L$.) Connections corresponding 
to
this differential extension of $A_{\theta}$ appear naturally in the 
study 
of toroidal compactifications of M(atrix) theory. 

  Let us suppose now that in addition to the action of Lie algebra on 
$A$
we have a finite group $G$ acting on $A$ and $L$ by means of 
automorphisms
and that actions of $G$ on $A$ and $L$ are compatible. (If we denote
automorphisms of $A$ and of $L$ corresponding to the element $\gamma\in 
G$
by the same letter $\gamma$ this means that $\gamma (T(a))=(\gamma 
T)\cdot
(\gamma a)$ for every $\gamma \in G,\ \ T\in L,\ \ a\in A$.) One can
define in natural way the action of $G$ on the algebra $\Omega =\Omega (L,A)$; this action
commutes with the differential. This means that we can regard the crossed product $\Omega\rtimes
G$ as a differential algebra; we have $(\Omega \rtimes G)_0=\Omega _0\rtimes G=A\rtimes G$ and
therefore the crossed product can be considered as differential extension of  $A\rtimes G$. 

  An $A\rtimes G$-module $E$ can be considered as an $A$-module 
equipped
with action of the group $G$ that is compatible with the action of $G$ 
on
$A$ (more precisely we should have $\gamma (xa)=\gamma (x)\cdot \gamma
(a)$). As always a connection on $E$ is defined as a connection 
$\nabla$
on $\Omega \rtimes G$-module
 $${\cal E}=E\otimes _{A\rtimes
G}(\Omega \rtimes G).$$ 
Again this connection is completely
determined by the map $\nabla : {\cal E}^0\rightarrow {\cal
E}^1$ that can be considered as a map
$$\nabla :E\rightarrow E\otimes _{A\rtimes G}(\Omega ^1(L,A)\rtimes 
G)$$
or as a map
$$\nabla :E\rightarrow E\otimes L^*$$
that determines a connection on $A$-module $E$ and is compatible with 
the
action of the group $G$ on $E$ and on $E\otimes L^*$.

  In the case when $A$ is an algebra of functions on noncommutative 
torus
the connections we obtained are precisely the connections that arise 
by
compactification of M(atrix) theory on toroidal orbifolds 
(see [9],[10]).

\centerline {{\bf  7. Deformations of  commutative algebras}}

The problem of deformation of algebras of functions on smooth manifolds 
is closely related to the problem of quantization.
It  can be formulated in the following way. Let $C(M)$ denote an  
algebra of functions on a smooth manifold $M$. To quantize a  manifold $M$ 
we should construct a family ${\cal A}_{\epsilon}$ of associative 
algebras obeying  ${\cal A}_{\epsilon =0}=C(M)$. This definition should be 
made more precise. First of all one can use various  algebras of 
functions on $M$ in this definition. If $M$ is compact it is natural to work 
with the algebra $C^{\infty}(M)$ of smooth functions on $M$, but in 
the  case of noncompact $M$ there are  various  interesting versions of 
$C(M)$. One can impose  various  conditions on dependence of  ${\cal 
A}_{\epsilon}$ from   $\epsilon$. In the most popular approach one 
considers   $\epsilon$ as a formal parameter; this means that  ${\cal 
A}_{\epsilon}$ coincides with  ${\cal A}_{\epsilon =0}=C^{\infty}(M)$ as a 
vector space and the product in  ${\cal A}_{\epsilon}$ (star-product) is 
a formal power series with respect to  $\epsilon$:
$$f\star g=f\cdot g+B_1(f,g)\epsilon +B_2(f,g)\epsilon ^2+... .$$
It is easy to check that the operation $\{ f,g\} =B_1(f,g)-B_1(g,f)$ 
(Poisson bracket corresponding to the family ${\cal A}_{\epsilon}$) 
specifies a structure of Poisson manifold on $M$, therefore usualy one 
speaks about quantization of Poisson manifolds. The problem of formal 
quantization was solved in the most important case of symplectic manifolds in 
[15] and for general Poisson manifolds in [16]. However, the situation in 
the case when ${\cal  A}_{\epsilon}$ is assumed to be a smooth family 
depending on parameter $\epsilon \in {\bf R}$ remains unclear. It is 
difficult to construct such families even for simple manifolds $M$. The 
most important constructions of this kind are based on the formula 
\begin {equation}
f\star g=\int \alpha _{\theta u}(f)\alpha _v(g)e^{iuv}du \  dv
\end {equation}
where $\alpha _u$ stands for a strongly continuous action of an abelian  
Lie group $L={\bf R}^d$ on associative algebra ${\cal A}_0$, $\theta$ 
is an antisymmetric $d\times d$ matrix (see [17] for details). The new 
product (star-product) determines an associative algebra ${\cal 
A}_{\theta}$ that depends continuously on $\theta$. 

Notice that in (20) we assumed that $u,v\in L={\bf R}^d$. It is more 
convenient to think that $v\in L,\  u\in L^*$ and $\theta$ is a linear 
operator acting from $L^*$ into $L$ and obeying $\theta ^+=-\theta$. This 
interpretation of (20) permits us to say that we don't need inner product 
on $L$ to apply (20). Applying (20) to  the algebra of smooth functions on 
torus $T^d={\bf R}^d /{\bf Z}^d$ with natural action of $L={\bf  R}^d$ 
we obtain a family ${\cal A}_{\theta}$ as a continuous deformation of 
the algebra ${\cal A}_0=C^{\infty}(T^d)$. This algebra is by 
definition an  algebra of smooth functions on noncommutative torus 
$T_{\theta}^d$. Analogously, we obtain various classes of functions on 
noncommutative euclidean space ${\bf R}_{\theta}^d$. For example we can fix $\rho 
\in (0,1]$ and denote by $\Gamma_{\rho}^m$ a class of smooth complex 
functions $a(x)$ on ${\bf R}^d$ obeying
$$|\partial _{\alpha}a(z)|\leq C_{\alpha}<z>^{m-\rho|\alpha|}$$
where $\alpha=(\alpha _1,...,\alpha _d),\  |\alpha|=\alpha 
_1+...+\alpha_d,\  m\in {\bf R},\  <z>=(1+|z|^2)^{1/2}$. Then one can prove that 
the star-product of  $a^{\prime}\in \Gamma_{\rho}^{m_1}$ and  
$a^{\prime\prime}\in \Gamma_{\rho}^{m_2}$ belongs to  $ \Gamma_{\rho}^{m_1+m_2}$. 
In particular,  $ \Gamma_{\rho}^m$ is an algebra for $m\leq 0$ and the 
union  $\Gamma_{\rho}$ of all  $\Gamma_{\rho}^m$  is also an algebra 
with respect to star-product. One more algebra can be obtained if we 
assume that $\theta ^{ij}=\epsilon \omega ^{ij}$ where $\epsilon$ is a 
formal parameter and consider star-product on the space $A^{\epsilon}$ of 
formal series 
\begin {equation}
a(x,\epsilon)=\sum _{k,l}\epsilon ^kP_l(x)
\end {equation}
where $k$ and $l$ are nonnegative integers and $p_l(x)$ stands for a 
polynomial of degree $l$  on ${\bf R}^d$. In all these cases we obtain an 
algebra ${\cal A}_{\theta}$ that represents a class of functions on 
noncommutative ${\bf R}^d$.  In particular,  we can assign an algebra 
${\cal A}({\cal T})$ to every  symplectic linear space ${\cal T}$. Now we 
can consider an arbitrary  symplectic manifold $M$ and a bundle of 
algebras ${\cal A}({\cal T}_x)$ corresponding to tangent spaces  ${\cal T}_x 
$ where $x\in M$. The set of  sections of this bundle also constitutes 
an associative algebra with respect to fibrewise multiplication. We 
shall denote this algebra by $W$ following Fedosov. It is necessary to 
emphasize that Fedosov considered only the case of the algebra of formal 
power series, but we use the  same notation $W$ in all cases. A part of 
Fedosov's constructions can be generalized immediately to other 
algebras. In particular, for every symplectic connection on $M$ we can 
construct an operator $\partial :W \otimes \Lambda\rightarrow W\otimes 
\Lambda$, acting on a tensor product of $W$ and the algebra $\Lambda$ of 
differential forms on $M$. In Darboux local coordinates $x^i$ this operator 
has the form 
\begin {equation}
\partial a =da+i[\Gamma,a]
\end {equation}
where 
$\Gamma = {1\over 2} \Gamma _{ijk}(x)y^iy^kdx^k,\ \  \Gamma _{ijk}$  is 
a symplectic tensor, and $y^i$ are coordinates on  tangent space ${\cal 
T}_x$. It is easy to check that 
\begin {equation}
\partial ^2a=[R,a]
\end {equation}
where 
\begin {equation}
R={a\over 4}R_{ijkl}y^iy^jdx^k\hat {d}x^l,\ \  
R_{ijkl}=\omega_{lm}R_{jkl}^m
\end {equation}
stands for the curvature tensor of  symplectic connection. (Our 
notations differ slightly from Fedosov's notations; to convince ourselves that 
formulas (23) and (24) coincide with the formulas in [18] we should take into 
account that Fedosov's $y^i$ contain an extra factor of $\epsilon 
^{1/2}$.) Let us suppose that the algebra ${\cal A}$ used in the 
construction of the algebra $W$ contains  polynomials ( for example we can take
 the algebra $\Gamma_{\rho}$ 
as  ${\cal A}$). Then $R$ belongs to  the algebra $W\otimes \Lambda$  
and we obtain the following statement.

  {\it The algebra $W\otimes \Lambda$ equipped with the operator 
$Q=\partial$ can be considered as a $Q$-algebra.}

The condition (1) follows from (23) and the condition (3) from the 
Bianchi identity $\partial R=0$.

Let us introduce an operator $Q_{\gamma}=\partial +\tilde {\gamma}$ 
where $\tilde {\gamma}x=[\gamma,x]$ and $\gamma\in W\otimes \Lambda ^1$.

 {\it The algebra $W\otimes \Lambda$ equipped with the operator 
$Q_{\gamma}$ is a $Q$-algebra; this $Q$-algebra is equivalent in the sense of 
Section 3 to the algebra $W\otimes \Lambda$ equipped with  operator 
$Q=\partial$.}

It follows from calculation of Sec. 3 that 
$$Q_{\gamma}^2=iR+\partial\gamma+\gamma^2.$$

We see that in the case when 
$$iR+\partial \gamma+\gamma ^2$$
is a central element we have $Q_{\gamma}^2=0$, i.e. 
 the our $Q$-algebra is a differential algebra that is  equivalent as a 
$Q$-algebra to $W\otimes \Lambda$ equipped with  $Q=\partial$. 
 
The Fedosov's approach to quantization of symplectic manifolds can be 
described as follows. In the framework  of formal power series with 
respect to $\epsilon$ we can find such a $\gamma=\gamma _0 $
that the operator $Q_{\gamma}$ is a  differential.

The elements of $W\subset W\otimes \Lambda$ that are annihilated by the 
operator $Q_{\gamma_0}$ form an associative algebra ${\cal 
A}_{\epsilon}$ that can be considered as a formal deformation of the algebra of 
functions on symplectic manifold $M$. The  associative algebra at hand is 
quasiisomorhic to  differential algebra $(W\otimes \Lambda, 
Q_{\gamma_0})$, i.e. to $W\otimes \Lambda$ equipped with  differential  
$Q_{\gamma_0}$. (M.Kontsevich, private communication). This means that modules 
over ${\cal A}_{\epsilon}$ are in one-to one correspondence with 
$dg$-modules over $(W\otimes \Lambda, Q_{\gamma_0})$ (for every ${\cal 
A}_{\epsilon}$-module one can construct a $dg$-module as tensor product of $E$ 
with $W\otimes \Lambda$ over ${\cal A}_{\epsilon}$; the  differential 
$Q_{\gamma_0}$ descends to a differential on this tensor product). From 
the other side there exists a one-to-one correspondence between 
$dg$-modules over  differential algebra $(W\otimes \Lambda, Q_{\gamma_0})$ and 
$dg$-modules over $Q$-algebra  $(W\otimes \Lambda, Q_{\gamma})$  for arbitrary  $\gamma \in
W\otimes 
\Lambda$. In particular, we can take $\gamma=0$ and study $dg$-modules 
over the $Q$-algebra  $(W\otimes \Lambda, \partial )$.

One interesting example of $dg$-modules over $(W\otimes \Lambda, \partial)$ can be constructed
in the case when $M$ is a Kaehler manifold and the symplectic connection $\partial $ corresponds
to the Kaehler metric on $M$. Then we can construct an $W$-module ${\cal F}$ as a module of
sections of the bundle of Fock modules. (Using Kaehler structure on $M$ we can define a Fock
representation of ${\cal A}({\cal T}_x)$ for every $x\in M$.)  The symplectic connection
$\partial $ acts on ${\cal F}\otimes \Lambda$; one can check that ${\cal F}\otimes \Lambda$ is a
$dg$-module over  $(W\otimes \Lambda, \partial)$ with respect to this action (this fact easily
follows from the results of [21]).

Notice that one can generalize  the above constructions, considering the case
when polynomials don't belong to the algebra $W$, but can be considered as
(left and right) multipliers on $W$. That we are dealing with generalized 
$Q$-algebras in the sense defined in Introduction; the module ${\cal F}\otimes \Lambda$ can be
equipped with zero curvature connection (is a $dg$-module) also in this more general case. 
 
\centerline {{\bf 8. Gauge theories on noncommutative tori.}}

   Let consider in more detail  modules over the algebra $A_{\theta}$  
of  smooth functions on $d$-dimensional noncommutative torus. Recall, 
that  $\theta$ stands for bilinear function on a lattice;
it will be convenient for us to identify it with  $d\times d$ matrix $ 
\theta^{ij}$.
Notice, that the algebra $A_{\theta}$ has a trace that is unique up to 
a constant factor ; it is given by the relations ${\rm Tr}U_n=0$ for 
$n\neq 0, \  {\rm Tr} U_0=1$. This trace induces a trace on the algebra 
End$_{A_{\theta}}E$ for every projective module $E$; the trace of unit 
endomorphism can be interpreted as (fractional) dimension of module 
$E$.

The algebra $A_{\theta}$ for $\theta =0$ (and, more generally, for 
integral $\theta$) is isomorphic to the algebra of smooth functions on 
usual torus. It depends continuously on parameter $\theta$, therefore one 
should expect that $K$-groups of $A_{\theta}$ coincide with $K$-groups 
of torus. This fact was proved in [19]. We can consider Chern character 
${\rm ch}E$  as an even element of Grassmann algebra $\Lambda$ having $d$ 
generators $\alpha ^1,...,\alpha ^d$. (By definition the Chern 
character is an element of $\bar {\Omega}_{\theta}=\Omega_{\theta}  
/[\Omega_{\theta} ,\Omega_{\theta} ]$. We can represent $\Omega_{\theta} $ as 
$A_{\theta} \otimes \Lambda$. The algebra $\Lambda$ is supercommutative, 
therefore $\bar {\Lambda}=\Lambda$. Uniqueness of trace on $A_{\theta}$ 
means that $\bar {A}_{\theta}={\bf C}$. We obtain $\bar 
{\Omega}_{\theta}=\Lambda$.) Operators of multiplication by $\alpha ^i$ and (left) 
derivatives $\partial /\partial\alpha^i$ satisfy canonical anticommutation 
relation (i.e. specify a representation of  Clifford algebra); this 
means that $\Lambda$ can be considered as a fermionic Fock space. One can 
prove, that 
$$\mu (E)=e^{{1\over 2}{\partial \over \partial 
\alpha^i}\theta^{ij}{\partial\over\partial\alpha^j}}{\rm ch}E$$
is an integral element of $\Lambda$ [19]. This element characterizes 
completely the $K$-theory class of   projective module $E$. One can identify 
the group $K_0 (E)$ with the lattice $\Lambda ^{{\rm even}}({\bf Z})$ of 
integral even elements of $\Lambda$.  The element $\mu (E)$ can be 
considered as a collection of topological numbers  corresponding to the 
module $E$. The group $K_1 (A_{\theta})$ can be identified with the 
lattice $\Lambda ^{{\rm  odd}}({\bf Z})$ of odd integral elements of 
$\Lambda$.
(Talking about  even and odd  elements of $\Lambda$ we have in mind 
Grassmann parity.)

If $E=E_0$ (i.e. graded module $E$ has only elements of degree $0$) 
then obviously  ${\rm dim}E={\rm Tr}1={\rm ch}_0E>0$. Expressing the 
zeroth component  ${\rm ch}_0E$ of Chern character ch$E$ in terms of 
$\mu(E)$ we obtain  necessary condition for existence of module obeying 
$E=E_0$ and having $\mu(E)=\mu$; such a module will be denoted by 
$E_{\mu}^{\theta}$ or simply by $E_{\mu}$ 
if $\theta$ is fixed. 

If  $\theta$ is irrational (i.e. has at least one irrational entry)  
then this condition is also sufficient [20]. For irrational $\theta$ two 
projective  modules having  only elements of degree $0$ are isomorphic 
iff they belong  to the same $K$-theory class [20]. This means that the 
module $E_{\mu}$ is unique (up to isomorphism). 

 For simplicity we'll assume that $\theta$ is irrational. Then every 
${\bf Z}_2$-graded projective module has a unique representation in the 
form $E_{\mu_1}+\Pi E_{\mu_2}$.

 Let $G$ be an abelian group that can be represented as a direct sum of  
${\bf R}^p$ and finite group. If $(\gamma, \tilde {\gamma})\in G\times 
G^*$ where  $G^*$ is the group of characters of $G$ one defines an
 operator  $U_{\gamma,\tilde {\gamma}}$ acting on functions on $G$  by the formula 

$$(U_{\gamma,\tilde {\gamma}}f)(x)=\tilde {\gamma}(x)f(x+\gamma).$$
More precisely, we should consider  $U_{\gamma,\tilde {\gamma}}$ as operators on the Schwartz
space ${\cal S}(G)$ (or the space of smooth functions on $G$ that tend to zero at infinity
faster than any power.)

If $\Gamma$ is a lattice in $G\times G^*$  (i.e. $\Gamma$ is a discrete 
subgroup of  $G\times G^*$ and  $G\times G^*/ \Gamma $ is compact )  
the operators  $U_{\gamma,\tilde {\gamma}}$ with   $(\gamma,\tilde 
{\gamma})\in \Gamma $ specify a projective module over  noncommutative torus 
; modules of  
such a kind are called Heisenberg modules [20]. Every Heisenberg module $E$ 
has a constant curvature connection; this means that ch$E$ is a 
quadratic exponent and $\mu (E)$ is a generalized  quadratic exponent  i.e. a 
limit of  quadratic exponents (see [5] for details). We'll say that a  
module  admitting constant curvature connections is a basic
module if it cannot  be represented as a direct sum of  isomorphic 
modules.  In other words a basic module is a module $E_{\mu}$ where $\mu$ 
is a generalized  quadratic exponent  and $\mu$  cannot be represented  
in the form  $k\mu _0$ where $\mu _0\in \Lambda ^{{\rm even}}({\bf 
Z}),\  k>1$. One can check that endomorphism algebra 
End$_{A_{\theta}}E_{\mu}^{\theta}$ of basic module $E_{\mu}^{\theta}$ is an algebra of 
functions  on another noncommutative torus. This algebra $A_{\hat {\theta}}$ 
is Morita equivalent to the original algebra $A_{\theta}$. One can 
consider a basic module $E_{\mu}^{\theta}$ as an $(A_{{\hat 
\theta}},A_{\theta })$-bimodule that establishes Morita equivalence between $A_{{\hat 
\theta}}$ and $A_{\theta }$.We'll denote this bimodule by  $E_{{\hat 
\theta},\theta}$. In particular, $E_{\theta ,\theta}=A_{\theta} $ where 
$A_{\theta}$  is considered as $(A_{\theta},A_{\theta})$-bimodule. 

One can check that $\hat {\theta}=g(\theta)$ where $g\in SO(d,d,{\bf Z})$ [5]. (The group
$SO(d,d,{\bf R})$ acts in the space of antisymmetric matrices by means of fractional  linear
transformations $g(\theta)=(A\theta +B)(C\theta +D)^{-1}$.)  More precise notation for  Morita
equivalence  bimodules $E_{\hat {\theta},\theta}$ should include the element $g\in SO(d,d,{\bf
Z})$ connecting $\theta$ and $\hat {\theta}$. The  bimodules $E_{\hat {\theta},\theta}$ can be
equipped with constant curvature connection. Using this connection we can establish gauge
Morita equivalence of $A_{\theta}$ and $A_{\hat {\theta}}$. One can prove that all gauge  Morita
equivalences  on noncommutative tori  are of this kind. In other words, tori $A_{\theta}$ and
$A_{\hat {\theta}}$ are gauge  Morita equivalent iff $\hat {\theta}=g(\theta)$ where $g\in
SO(d,d,{\bf Z})$ [5].  

It is obvious that the tensor product of $E_{\hat {\theta},\theta}$ and $E_{\theta,\theta
^{\prime}}$ over $A_{\theta}$ is a bimodule that establishes  Morita equivalence between  
$A_{\hat {\theta}}$ and $A_{\theta ^{\prime}}$; we will write 
\begin {equation}
E_{\hat {\theta},\theta}\otimes _{A_{\theta}}E_ {\theta,\theta ^{\prime}}=E_{\hat
{\theta},\theta ^{\prime}}
\end {equation}

This means, in particular, that one can define natural   
$E_{\hat {\theta},\theta^{\prime}}$-valued pairing between    $E_{\hat{\theta},\theta}$ and   
$E_{\theta,\theta^{\prime}}$.

Let us describe the space $\{ E_{\mu_1}\rightarrow E_{\mu_2}\} $ of 
$A_{\theta}$-linear maps of basic   $A_{\theta}$-module $E_{\mu_1}$ into 
basic $A_{\theta}$-module $E_{\mu_2}$. We can consider $E_{\mu_1}$ as 
$(A_{\theta_1},A_{\theta})$-bimodule  $E_{\theta_1,\theta}$   and 
$E_{\mu_2}$  as $(A_{\theta_2},A_{\theta})$-bimodule   $E_{\theta_2,\theta}$.  
The space $\{ E_{\mu_1}\rightarrow E_{\mu_2}\} $ can be considered  as 
$(A_{\theta_2},A_{\theta_1})$-bimodule. One can check {\it this bimodule 
establishes Morita equivalence between} $A_{\theta_2}$ {\it and} $A_{\theta _1}$; 
this permits us to identify it with $E_{\theta_2,\theta_1}$.
  
     Let us consider now matrices $\theta_1,...,\theta_k$ belonging to 
the same orbit of  SO$(d,d,{\bf Z})$ in the space of antisymmetric 
$d\times d$ matrices. Corresponding noncommutative tori are Morita 
equivalent,  therefore we can consider bimodules $E_{\theta_i,\theta_j}$. Let 
us consider a space   $A_{\theta_1,...,\theta_k}$ consisting of $k\times 
k$ matrices where the entry in the $i$-th row and $j$-th column belongs 
to  $E_{\theta_i,\theta_j}$. It follows from (25) that there exists 
natural pairing between  $E_{\theta_i,\theta_j}$ and  
$E_{\theta_j,\theta_l}$ with values in  $E_{\theta_i,\theta_l}$. Using this pairing we define 
an algebra structure on  $A_{\theta_1,...,\theta_k}$.

  It follows from the results mentioned above that every algebra 
$A_{\theta_1,...,\theta_k}$ is  Morita equivalent to noncommutative torus 
and, conversely, every algebra that is Morita equivalent to  
noncommutative torus is isomorphic to one of  algebras $A_{\theta_1,...,\theta_k}$. 
(Recall, that we assume that the parameter of  noncommutativity  
$\theta$ is an irrational matrix.) 

The proof  is based on identification of an algebra Morita equivalent 
to $A_{\theta}$ with ${\rm End}_{A_{\theta}}E$ where $E$ is a projective 
$A_{\theta}$-module and on remark that $E$ can be represented as a 
direct sum of basic modules [20].(We used the fact that every projective 
$A_{\theta}$-module is a generator and therefore can be used to construct Morita equivalence
between $A_{\theta}$ and End$_{A_{\theta}}E $.) It follows from the above statements that the
endomorphism algebra of a direct sum of basic modules can be
considered as algebra  $A_{\theta_1,...,\theta_k}$ and that
every algebra  $A_{\theta_1,...,\theta_k}$ can be obtained this way.

Notice, that there exist numerous non-trivial isomorphisms between  
algebras $A_{\theta_1,...,\theta_k}$. This follows from the fact that the 
relation $\mu_1+...+\mu_k=\mu_1^{\prime}+...+\mu_l^{\prime}$ implies  an 
isomorphism of $A_{\theta}$-modules $E_{\mu_1}+...+E_{\mu_k}$ and 
$E_{\mu_1^{\prime}}+...+E_{\mu_k^{\prime}}$ (here $E_{\mu_i}$  and 
$E_{\mu_j^{\prime}}$ are basic  $A_{\theta}$-modules  obeying $\mu 
(E_{\mu_i})=\mu_i ,\ \  \mu (E_{\mu_j^{\prime}})=\mu_j^{\prime}$). 
Isomorphisms of corresponding endomorphism  algebras leads to a 
conclusion that  $A_{\theta_1,...,\theta_k}$ is isomorphic to 
$A_{\theta_1^{\prime},...,\theta_l^{\prime}}$  where $\theta _i$ and 
$\theta_j^{\prime}$ are defined by the formula
$$A_{\theta_i}={\rm End}_{A_{\theta}}E_{\mu_i},\ \ \   
A_{\theta_j^{\prime}}={\rm End}_{A_{\theta}}E_{\mu_j^{\prime}}.$$
This result can be reformulated in the following way. Let us suppose 
that $\theta _i=g_i(\theta),\  \theta_j^{\prime}=g_j^{\prime}(\theta)$  
and  $\sum g_i(1)=\sum g_j^{\prime}(1)$ where $g_i,\  g_j^{\prime}$  are 
elements of the group $SO(d,d,{\bf Z})$ acting on $\theta$ by means of 
fractional linear transformations and on elements of  $\Lambda $  (on Fock 
space)  by means of linear canonical transformations (=spinor 
representation).  Then $A_{\theta_1,...,\theta_k}$ is isomorphic to 
$A_{\theta_1^{\prime},...,\theta_l^{\prime}}$. 

 In the dimensions $d=2,3$ every projective module is isomorphic to the direct sum of identical
basic modules. (This follows from the fact that in this dimensions $\mu(E)$ is always a
generalized quadratic exponent, hence every module admits a constant curvature connection). We
obtain that in the dimensions $2,3$ every algebra that is  Morita equivalent to $A_{\theta}$ is
isomorphic to matrix algebra Mat$_n(A_{\hat {\theta}})$ where $\hat {\theta}=g(\theta),\ \  g\in
SO(d,d,{\bf Z})$.

Notice, that bimodules $E_{\hat {\theta},\theta}$  and algebras $A_{\theta_1,...,\theta_k}$ have
a nice physical interpretation. It was shown in [8] that  $E_{\hat {\theta},\theta}$ can be
interpreted as a state space of strings connecting two $D$-branes carrying gauge theories
corresponding to noncommutative tori $A_{\hat {\theta}}, A_{\theta}$. One can say that
$A_{\theta_1,...,\theta_k}$ is an algebra of string states in presence of $k$ $D$-branes. (It
was shown in [8] that different boundary conditions for given theory in the bulk correspond to
Morita equivalent algebras. In our case we have $k$ Morita equivalent tori.)

A connection on $A_{\theta}$-module $E$ can be identified with a connection on $\Omega
_{\theta}$-module   $E\otimes _{A_{\theta}}\Omega _{\theta}$ where  $\Omega _{\theta}=\Omega
(L,A_{\theta})$ denotes a differential extension of $A_{\theta}$ corresponding abelian Lie
algebra $L$ of derivations of $A_{\theta}$. Gauge  Morita equivalence of $A_{\theta}$ and
$A_{\hat {\theta}}$ implies  Morita equivalence of $Q$-algebras $\Omega _{\theta}\  \Omega
_{\hat {\theta}}$ and physical   equivalence of corresponding gauge theories. We can generalize
this statement considering $Q$-extension $\Omega _{\theta_1,...,\theta_k}$ of the algebra
$A_{\theta_1,...,\theta_k}$. The definition of the space  $\Omega _{\theta_1,...,\theta_k}$
repeats the definition of   
 $A_{\theta_1,...,\theta_k}$, but instead of bimodules $E_{\theta_i,\theta_j}$ we should use
bimodules ${\cal E}_{\theta_i,\theta_j}$ that establish  Morita equivalence between  $\Omega
_{\theta_i}$ and $\Omega _{\theta_j}$. Using the natural  $\Omega _{\theta_i,\theta_l}$-valued
pairing between ${\cal E}_{\theta_i,\theta_j}$ and ${\cal E}_{\theta_j,\theta_l}$ we introduce
an algebra structure on  $\Omega _{\theta_1,...,\theta_k}$. One can consider  $\Omega
_{\theta_1,...,\theta_k}$ as a $Q$-algebra introducing $Q$ as an operator acting on every
bimodule ${\cal E}_{\theta_i,\theta_j}$ as constant curvature connection. (Such a connection is
determined uniquely up to an additive constant.) It is easy to check that $Q^2x=[\omega,x]$,
where $\omega=$diag$(\omega_1,...,\omega_k)$, where $\omega_i\in {\cal
E}_{\theta_i,\theta_i}=\Omega _{\theta_i}$ is a scalar (an element of the form const$\cdot
1$). It follows from the above consideration that  $\Omega _{\theta_1,...,\theta_k}$  is  Morita
equivalent to $\Omega_{\theta_i}$ as a $Q$-algebra. This means that {\it the gauge theory based
on the} $Q$-{\it algebra} $\Omega _{\theta_1,...,\theta_k}$ {\it is physically equivalent to the
gauge theory on noncommutative torus.}

In the case of two-dimensional noncommutative tori we can make the consideration above more
explicit using the results of [22] (see also [23] for more complete formulas and some
applications). In this case we can work with the group $SL(2,{\bf Z})$  instead of $SO(2,2,{\bf
Z})$.  Representing a $2\times 2$ antisymmetric matrix as 
\[  \left(  \begin {array} {cc}
0 & \theta \\
-\theta & 0 \\
\end {array}  \right)  \]  
we consider the noncommutativity parameter as a real number $\theta$;  the group $SL(2,{\bf Z})$
acts on $\theta$ by means of fractional linear transformations. A basic bimodule $E_{\hat
{\theta},  \theta}$ can be realized in the space  ${\cal S}({\bf R}\times {\bf Z}_m)$ by means
of operators 
\begin {equation}
(U_1 f)(x,\mu)=f(x-r,\mu -\rho),
\end {equation}
\begin {equation}
(U_2 f)(x,\mu)=\exp [2\pi  i (tx-s\mu )]  f(x,\mu ),
\end {equation}
\begin {equation}
(Z_1 f)(x,\mu)=f(x-r^{\prime},\mu -\rho ^{\prime}),
\end {equation}
\begin {equation}
(Z_2 f)(x,\mu)=\exp [2\pi i (t^{\prime}x-s^{\prime}\mu)]  f(x,\mu ).
\end {equation}
Here $\hat {\theta}={b+a\theta \over  n+m\theta},\ \  x\in {\bf R},\ \  \mu \in {\bf Z}_m$, the
numbers $r,\  t,\  s,\  r,^{\prime}\  t^{\prime},\  s^{\prime}\in {\bf R}$ and $\rho,\
\rho^{\prime}\in {\bf Z}_m$  are chosen in such a way that the operators $U_1,U_2$ obey
$U_1U_2=e^{-2\pi i\theta}U_2U_1$ and therefore specify a right $A_{\theta}$-module
$E_{n,m}(\theta)$, the operators  $Z_1,Z_2$, commuting with $U_1,U_2$, obey $Z_1Z_2=e^{2\pi
i\hat {\theta}}Z_2Z_1$ and specify a left $A_{\hat {\theta}}$-module. 

We take $r={1\over m},\  \rho=-n,\  t=n+m\theta,\  s=-{1\over m},\  r^{\prime}={1\over
m(n+m\theta)},\  \rho^{\prime}=-1,\  t^{\prime}=1,\  s^{\prime}=-{a\over m}$. 

The natural bilinear $E_{\hat {\theta},\theta^{\prime}}$-valued pairing between $E_{\hat
{\theta},\theta}$ and $E_{\theta, \theta^{\prime}}$ sends a pair $(f,g)$ into \begin {equation}
h(x,\Delta)=\sum_{q\in {\bf Z}}f(Ax+Bq+C\Delta, Dq+E)  g(\tilde{A}x+\tilde{B}q+\tilde{C}\Delta,
\tilde {D}q+\tilde {E}).
\end {equation}
Here $\hat {\theta}=(b+a\theta )(n+m\theta)^{-1},\ \
\theta=(\beta+\alpha\theta^{\prime})(k+l\theta^{\prime})^{-1}$, the bimodule $E_{\hat
{\theta},\theta^{\prime}}$ consists of functions of $x\in {\bf R},\  \Delta  \in {\bf
Z}_{nl+m\alpha}$ (it is isomorphic to $E_{bk+a\alpha , nl+m\alpha}(\theta ^{\prime})$ as
$A_{\theta ^{\prime}}$-module).  The coefficients in (30) are given by the formulas:  

$$A=1,\  B={1\over m},\  C=-{l\over m(nl+m\alpha)},\  D=-n,\  E=1,$$

$$\tilde {A}=n+m\theta,\  \tilde {B}={l\theta-\alpha\over l},\  \tilde {C}={\alpha-l\theta\over
nl+m\alpha},\  \tilde {D}=1,\  \tilde {E}=0.$$

Remark. Our considerations were not completely rigorous in the following relation. The bilinear
$E_{\hat {\theta},\theta }$-valued pairing,  $\rho_{\hat
{\theta},\theta|\theta,\theta^{\prime}}^{\hat {\theta},\theta ^{\prime}}$ between $E_{\hat
{\theta},\theta}$ and $E_{\theta,\theta^{\prime}}$ is not completely canonical; it is defined
only up to a constant factor. (Multiplication by a constant can be considered as an automorphism
of a bimodule.) This means that the idenfication (25) is possible, but not unique. In principle,
associativity of tensor product can be violated:
\begin {equation}
\rho_{\hat {\theta},\theta^{\prime}|\theta^{\prime},\theta^{\prime \prime}}^{\hat
{\theta},\theta ^{\prime \prime}}(\rho_{\hat {\theta},\theta |\theta,\theta^{\prime}}^{\hat
{\theta},\theta ^{\prime}}\otimes  1_{\theta^{\prime},\theta_{\prime \prime}})={\rm const}\cdot
\rho_{\hat {\theta},\theta |\theta,\theta^{\prime\prime}}^{\hat {\theta},\theta ^{\prime
\prime}}(1_{\hat {\theta},\theta}\otimes
\rho_{\theta,\theta^{\prime}|\theta^{\prime},\theta^{\prime\prime}}^{\theta,\theta 	^{\prime
\prime}}), 
\end {equation}
where  the constant is not necessarily equal to $1$. It is easy to see that this problem does
not appear in two-dimensional case: defining the pairing by (30) we obtain (31) with const$=1$
(see [23] for detailed calculation). This statement answers a question asked by Yu. Manin in
[24]. Moreover in any  dimension the constant in (31) is equal to $1$  for the appropriate
choice of pairing. ( We implicitly assume that such a choice is made; only with such a choice we
can say that the algebra $A_{\theta _1,...,\theta _k}$ is associative.) The absence of
appropriate choice would lead to unremovable non-associativity of the algebra $A_{\theta
_1,...,\theta _k}$. From the other side we know that an associative algebra End$_{A_{\theta}}E$
of endomorphisms of direct sum$E$ of basic modules can be interpreted as  $A_{\theta
_1,...,\theta _k}$. 

\centerline {{\bf Conclusion.}}
  
  In present paper we generalized the theory of connections on modules
over associative algebra. We embedded this theory into the theory of  
connections on modules over $Q$-algebras and proved a general duality
theorem in this framework. Namely, we proved that under certain 
conditions
there exists one-to-one correspondence between connections on modules 
over
one $Q$-algebra and connections on modules over another $Q$-algebra 
and
found
relation between corresponding curvatures. More precisely, it follows
from our results that  under certain conditions gauge theory 
constructed
by means of  
$Q$-algebra $\Omega$ is equivalent to gauge theory corresponding to 
the
$Q$-algebra $\rm End_{\Omega}\cal E$ where $\cal E$ is an
$\Omega$-module equipped with a connection.( We use the fact that every 
connection determines a structure of $Q$-algebra on the algebra of
endomorphisms.) This theorem can be applied to many concrete
situations; we gave an example in Sec. 8. 

We have shown that $Q$-algebras appear naturally in Fedosov's
quantization of symplectic manifolds and conjectured that they
can be used to circumvent the problems arising in the attempts
to quantize symplectic manifolds beyond the framework of 
perturbation theory.

\centerline {{\bf Acknowledgments.}}

I am very grateful to A. Connes, M. Khovanov, M. Kontsevich, Yu. Manin, A. 
Polishchuk, L. Positselsky, M. Rieffel and D. Sternheimer for useful 
comments.
I am deeply indebted to Caltech Theory Group, IHES and ESI for warm hospitality .
   
\centerline {{\bf References.}}

  1. Connes, A., {\it Noncommutative Geometry}, Academic Press,New 
York,(1994).

  2. Karoubi, M., {\it Homology cyclique et K-theorie}, Asterisque, 
149 (1987), 147 pp.

  3. Kastler, D., {\it Cyclic cohomology within the differential 
envelope}, Hermann, Paris, (1988),184 pp.

  4. Connes, A., Douglas, M., and Schwarz, A., {\it Noncommutative
geometry and Matrix theory: compactification on tori}, 
JHEP {\bf 02} (1998) 003; hep-th/9711162.

  5. Schwarz, A., {\it Morita equivalence and duality}, Nucl. Phys. 
{\bf B534} (1998) pp. 720-738; hep-th/9805034.

  6. Konechny, A. and Schwarz, A., {\it BPS states on noncommutative 
tori and duality}, Nucl. Phys. {\bf B550} (1999) 561-584; hep-th/9811159.

\ \ \ \ \   Konechny, A. and Schwarz, A.,{\it Supersymmetry algebra
and BPS states of super Yang-Mills theories on noncommutative tori}, 
Phys. Lett. {\bf B453} (1999) 23-29; hep-th/9901077.

\ \ \ \ \   Konechny, A. and Schwarz, A.,{\it 1/4 BPS states on noncommutative tori}, JHEP {\bf
09} (1999) 030; hep-th/9907008.

  7. Nekrasov, N. and Schwarz, A., {\it Instantons on noncommutative
$R^4$, and $(2,0)$ superconformal six-dimensional theory}, 
hep-th/9802068.

  8. Seiberg, N. and Witten, E., {\it String Theory and Noncommutative
Geometry}, JHEP {\bf 9909} (1999) 032; hep-th/9908142.

  9. Ho, P.-M. and Wu, Y.-S., {\it Noncommutative
Gauge Theories in Matrix Theory}, 
Phys.Rev. {\bf D58} (1998) 066003; hep-th/9801147.

 10. Konechny, A. and Schwarz, A.,{\it Compactification of M(atrix) theory on noncommutative
toroidal orbifolds}, hep-th/9912185

 11. Schwarz, A., {\it Noncommutative supergeometry and duality}, 
lett. Math. Phys.{\bf 50} (1999) 309-321; hep-th/9912212.  

 12. Bass, H., {\it Algebraic K-theory}, Benjamin, NY-Amsterdam, 
(1962), 762 pp

 13. Schwarz, A., 
{\it Geometry of Batalin-Vilkovisky quantization}, CMP, 
{\bf 155} (1993) 249-260,

\ \ \ \ \  {\it Semiclassical approximation in
Batalin-Vilkovisky formalism}, CMP,{\bf 158} (1993) 373-396,

\ \ \ \ \  Alexandrov,M., Kontsevich, M., Schwarz, A. and Zaboronsky, O.
,{\it  The geometry of master eqution and topological quantum field
theory}, Int. J. of Modern Physics,{\bf  A12} (1997) 1405-1429 

  14. Positselsky, L., {\it Nonhomogeneous quadratic duality and curvature},
Funk. Anal. Appl. {\bf 27} (1993) 197-204

  15. Bayen,F., Flato, M., Fronsdal,C., Lichnerowicz, A. and Sternheimer, D., {\it Deformation
theory and quantization I}, Ann of Phys. {\bf 111} (1978) 61-110

\ \ \ \ \  De Wilde, M. and Lecomte, P.B., {\it Existence of star-products and of formal
deformations of the Poisson Lie algebra of  arbitrary symplectic manifolds},
Lett. Math. Phys. {\it 7} (1983) 487-496       

  16. Kontsevich, M., {\it Deformation quantization of Poisson manifolds}, q-alg/9709040

  17. Rieffel, M., {\it Deformation and quantization for actions of} ${\bf R}^d$, Memoirs of
AMS,{\bf 106(506)} (1993) 1-93
  
  18. Fedosov, B. V., {\it A simple geometrical construction of deformation quantization},
J. Differential Geom. {\bf 40} (1994) 213-238

   19. Elliott, G.A., {\it On the K-theory of the $C^{\ast}$-algebra 
generated  by a projective representation of a torsion-free discrete 
abelian group} in {\it Operator Algebras and Group  representations}, 
Pilman, London, {\bf Vol.1} (1984) 159-164.

  20. Rieffel, M., {\it Projective modules over higher-dimensional 
noncommutative tori}, Can. J. Math. {\bf Vol.XL, No 2} {1999} 257-338
      
     {\it Noncommutative tori-a case study of noncommutative 
differentiable manifolds}, Contemp. Math. {\bf 105} (1990) 191-211.
 
   21. Karabegov, A.V., {\it On Fedosov's approach to deformation  quantization with separation
of variables}, QA/9903031 (1999)
 
   22. Dieng, M., and Schwarz, A., {\it Differential and complex geometry of two-dimensional
noncommutative tori}, QA/0203160 (2002).

   23. Polishchuk, A. and Schwarz, A.,{\it Categories of holomorphic vector bundles over
two-dimensional noncommutative tori}, (in preparation)

   24. Manin, Yu., {\it Real Multiplication and noncommutative geometry}, AG/0202109 (2002)
\end{document}